\documentclass[twocolumn,secnumarabic,amssymb, nobibnotes, aps, prl]{revtex4-1}
\usepackage{mathrsfs}
\begin{document}

\title{Tolman mass, generalized surface gravity, and entropy bounds}%

\author{Gabriel Abreu}%
\email{gabriel.abreu@msor.vuw.ac.nz}
\affiliation{School of Mathematics, Statistics, and Operations Research; Victoria University of Wellington, Wellington, New Zealand.}
\author{Matt Visser}%
\email{matt.visser@msor.vuw.ac.nz}
\affiliation{School of Mathematics, Statistics, and Operations Research; Victoria University of Wellington, Wellington, New Zealand.}

\date{7 May 2010; 25 May 2010; 28 June 2010; \LaTeX-ed \today}%

\begin{abstract}
In any static spacetime the quasi-local Tolman mass contained within a volume can be reduced to a Gauss-like surface integral involving the flux of a suitably defined generalized surface gravity. By introducing some basic thermodynamics and invoking the Unruh effect one can then develop elementary  bounds on the quasi-local entropy that are very similar in spirit to the holographic bound, and closely related to entanglement entropy.  
\end{abstract}


%-----------------------------------------------------------------------------------------------------------------------------------------
\maketitle
%-----------------------------------------------------------------------------------------------------------------------------------------
\def\sech{{\mathrm{sech}}}
\def\ln{{\mathrm{ln}}}
\def\d{{\mathrm{d}}}
\def\tr{{\mathrm{tr}}}
\def\A{{\mathscr{A}}}
%-----------------------------------------------------------------------------------------------------------------------------------------
%------------------------------------------------------------------------------------------------------------------------------------------
% very standard definitions
%------------------------------------------------------------------------------------------------------------------------------------------
\def\d{{\mathrm{d}}}
\newcommand{\scri}{\mathscr{I}}
\newcommand{\sun}{\ensuremath{\odot}}
\def\J{{\mathscr{J}}}
\def\sech{{\mathrm{sech}}}

%------------------------------------------------------------------------------------------------------------------------------------------
%------------------------------------------------------------------------------------------------------------------------------------------
%------------------------------------------------------------------------------------------------------------------------------------------
%------------------------------------------------------------------------------------------------------------------------------------------
%------------------------------------------------------------------------------------------------------------------------------------------
%------------------------------------------------------------------------------------------------------------------------------------------

Tolman mass is one of the standard notions of quasi-local mass in common use in general relativity \cite{Tolman0}. 
Using classical general relativity, this quasi-local Tolman mass can, in any static spacetime (either with or without a black hole region),  be reduced to a flux integral of (generalized) surface gravity across the boundary of the region of interest. (This
is closely related to the classical laws of black hole mechanics~\cite{Mechanix}.) General relativistic thermodynamics, together with a minimal appeal to quantum physics as embodied in the Unruh effect~\cite{notes},  is then sufficient to develop elementary but powerful bounds on a suitably defined notion of quasi-local entropy --- bounds very similar in spirit to the holographic bound~\cite{holographic, Bekenstein-gsl, Bekenstein-bound, paddy1, paddy2}, and closely related to entanglement entropy~\cite{Srednicki}.

In a static spacetime where the metric is taken to be of the form 
\begin{equation}
\label{E:1}
\d s^2 = - e^{-2\Psi} \, \d t^2 + g_{ij} \; \d x^i \d x^j,
\end{equation}
the Tolman mass contained in a region $\Omega$ is defined in terms of the orthonormal components of stress-energy by first taking
$\rho = T_{\hat 0 \hat 0}$ and  $p = {1\over3}  \tr\{T_{\hat \imath\hat \jmath} \}$; 
and then setting
\begin{equation}
m_T(\Omega) = \int_\Omega \sqrt{-g_4} \left\{ \rho + 3 p \right\} \d^3 x.
\end{equation}
The Einstein equations then imply the purely geometrical statement
\begin{equation}
m_T(\Omega) = {1\over4\pi} \int_\Omega \sqrt{-g_4} \; R_{\hat 0\hat 0} \; \d^3 x.
\end{equation}
The Tolman mass is intimately related to the Komar mass~\cite{Komar}, though we will not be phrasing any of the discussion below in terms of Killing vectors. 
It is a very old result, going back at least as far as Landau--Lifshitz~\cite{Landau} that in any stationary metric
\begin{equation}
R_{0}{}^{0} = {1\over\sqrt{-g_4}} \, \partial_i \left( \sqrt{-g_4} \, g^{0a} \, \Gamma^i{}_{a0} \right).
\end{equation}
(Here $a\in\{0,1,2,3\}$; $i\in\{1,2,3\}$.)
Adopting the manifest static coordinates of equation (\ref{E:1}), and then going to an orthonormal basis, this is more simply phrased as 
\begin{equation}
R_{\hat 0\hat 0} = {1\over\sqrt{-g_4}} \, \partial_i \left(\sqrt{-g_4} \, g^{00} \, \Gamma^i{}_{00} \right). 
\end{equation}
To get a clean physical interpretation of this formula, consider a fiducial observer (FIDO) with 4-velocity 
\begin{equation}
V^a = \left( \sqrt{|g^{00}|}; \; 0,0,0 \right).
\end{equation}
By definition the 4-acceleration of these FIDOs is
\begin{eqnarray}
A^a &=& (\nabla_V V)^a =  V^b \nabla_b V^a =  V^0 \left( \partial_0 V^a + \Gamma^{a}{}_{c0} V^c \right) 
\nonumber\\
&=& \sqrt{|g^{00}|} \, \Gamma^{a}{}_{00} \, \sqrt{|g^{00}|} =  |g^{00}|  \, \Gamma^{a}{}_{00}. 
\end{eqnarray}
But then, since $V$ is 4-orthogonal to $A$, we have
\begin{equation}
A^0=0;   \qquad A^i =  |g^{00}| \,  \Gamma^{i}{}_{00}; 
\end{equation}
where $A^i$ are the 3 spatial components of 4-acceleration. 
Therefore in any static spacetime, in the region outside the horizon,  the Landau--Lifshitz result is
\begin{equation}
R_{\hat 0\hat 0}  =  {1\over\sqrt{-g_4}} \, \partial_i \left(\sqrt{-g_4} A^i \right).
\end{equation}
Then for any 3-volume $\Omega$ (if a horizon is present then for convenience we confine ourselves to a region that lies outside the horizon) we can use ordinary partial derivative integration by parts to deduce 
\begin{eqnarray}
\int_\Omega \sqrt{-g_4} \, R_{\hat 0\hat 0} \, \d^3 x  &=& \int_\Omega  \partial_i \left(\sqrt{-g_4} \, A^i \right)  \d^3 x 
\nonumber\\
&=& \int_{\Omega} \partial_i \left( \sqrt{g_3} \, \{e^{-\Psi} A^i\} \right)  \, \d^3 x  
\nonumber\\
&=& \int_{\partial\Omega}  \{e^{-\Psi} A^i\}  \; \hat n_i \; \sqrt{g_2}\,  \d^2 x, 
\end{eqnarray}
where $\hat n$ is the unit normal,  (defined in terms of the 3-metric $g_{ij}$), and  $\sqrt{g_2}$ is the induced area measure on $\partial\Omega$.
Define a (generalized) surface gravity (3-vector) and its norm by
\begin{equation}
\kappa^i = e^{-\Psi} A^i; \qquad   \kappa = \sqrt{g_{ij} \, \kappa^i \kappa^j} = e^{-\Psi}  \sqrt{g_{ij} \, A^i A^j}.
\end{equation}
This is just the ``redshifted'' 4-acceleration of the FIDOs, and is a natural generalization of surface gravity, not just for any event horizon that might be present, but also applying to FIDOs skimming along the boundary $\partial\Omega$. In terms of this generalized surface gravity we now have
\begin{equation}
m_T(\Omega) = {1\over4\pi} \int_{\partial\Omega}  \kappa^i  \; \hat n_i \; \sqrt{g_2} \, \d^2 x = {1\over4\pi}  \int_{\partial\Omega}  \vec\kappa\cdot\hat n \; \d \A.
\end{equation}

Defining an average surface gravity $\bar \kappa(\partial\Omega)$, a total area $\A(\partial\Omega)$, and temporarily assuming we have no black hole regions to deal with, we see
\begin{equation}
m_T(\Omega) 
 \leq  {1\over 4\pi} \int_{\partial\Omega} \kappa \; \d(\mathrm{area}) = {\bar\kappa(\partial\Omega) \times \A(\partial\Omega)\over4\pi}.
\end{equation}
So for any (static) horizonless object such as a star or planet, (or monster~\cite{monster}, or gravastar~\cite{gravastar}, or black star~\cite{black-star}, or quasi-black hole~\cite{qbh}), we can bound its total Tolman mass in terms quantities measurable on its surface:
\begin{equation}
m_T(\Omega) \leq {\bar\kappa(\partial\Omega) \times  \A(\partial\Omega)
\over4\pi}.
\end{equation}
This  gives us a very general bound on the Tolman mass.

%------------------------------------------------------------------------------------------------------------------------------------------
%------------------------------------------------------------------------------------------------------------------------------------------

Now introduce thermodynamics: Consider the Euler relation (essentially the Gibbs--Duhem relation) for the entropy density of matter --- we are thinking of some equilibrium collection of atoms/\-molecules/\-fields making up a star/\-planet/\-monster/\-gravastar/black~star/ quasi-black hole. (No event horizons for now.) Then
\begin{equation}
s = {\rho + p - \mu n\over T},
\end{equation}
where (as previously) $p = {1\over3}\mathrm{tr}\{T_{\hat \imath\hat \jmath}\}$. 
The total entropy due to matter inside any specified 3-volume is
\begin{equation}
S(\Omega) = \int_\Omega \sqrt{g_3} \; s \; \d^3 x = \int_\Omega \sqrt{g_3} \; {\rho + p - \mu n\over T} \, \d^3 x.
\end{equation}
But the Tolman equilibrium conditions are~\cite{Tolman0, Tolman1, Tolman2, Balazs}
\begin{equation}
T \sqrt{-g_{00}} = T_\infty; \qquad \mu \sqrt{-g_{00}} = \mu_\infty;
\end{equation}
where we assume asymptotic flatness and without loss of generality set $g_{00}\to 1$ at spatial infinity. 
Then
\begin{equation}
S(\Omega)  = {1\over T_\infty} \int_\Omega \sqrt{-g_4} \, \{\rho + p\} \, \d^3 x - {\mu_\infty\over T_\infty} \, \int_\Omega \sqrt{g_3} \,n \,\d^3 x,
\end{equation}
that is
\begin{equation}
S(\Omega)  = {1\over T_\infty} \int_\Omega \sqrt{-g_4}  \,\{\rho + p\} \,\d^3 x - {\mu_\infty \, N\over T_\infty}.
\end{equation}
But thermodynamic stability requires $\mu \geq 0$, so
\begin{equation}
S(\Omega)  \leq {1\over T_\infty} \int_\Omega \sqrt{-g_4} \,  \{\rho + p\} \, \d^3 x. 
\end{equation}
Furthermore, in any system such as a star or planet $p>0$ throughout the interior so we have
\begin{equation}
S(\Omega)  \leq {1\over T_\infty} \int_\Omega \sqrt{-g_4}  \, \{\rho + 3p\} \, \d^3 x,
\end{equation}
which implies
\begin{equation}
S(\Omega)  \leq {m_T(\Omega)\over T_\infty}.
\end{equation}
That is --- the entropy inside any equilibrium star/planet  or monster/\-gravastar/black star/quasi-black hole is bou\-nd\-ed by the Tolman mass divided by the temperature (normalized at infinity).
By our theorem above
\begin{equation}
S(\Omega)  \leq  {\bar\kappa(\partial\Omega) \times  \A(\partial\Omega)
\over4\pi  T_\infty},
\end{equation}
where so far we have only used basic thermodynamics and no curved space quantum field theory.

Furthermore, due to the existence of the Unruh acceleration radiation phenomenon~\cite{notes}, we can argue that an observer at position $x$ on the boundary $\partial\Omega$ will see a \emph{minimum} locally measured temperature of
\begin{equation}
T(x) \geq T_\mathrm{Unruh}(x) =  {||A(x)||\over 2\pi}, 
\end{equation}
which when redshifted to infinity implies
\begin{equation}
T_\infty\geq \max_{x\in\partial\Omega} \left\{ \sqrt{-g_{00}(x)} \; T_\mathrm{Unruh}(x) \right\} =  \max_{x\in\partial\Omega}\left\{  {\kappa(x) \over 2\pi} \right\}.
\end{equation}
So the equilibrium temperature of a star/planet/ monster/gravastar/black~star/quasi-black hole confined inside a boundary $\partial\Omega$ must satisfy
\begin{equation}
T_\infty  \geq {\bar \kappa(\partial\Omega) \over 2\pi}.
\end{equation}
So finally
\begin{equation}
S(\Omega) \leq {\A(\partial\Omega)\over 2}.
\end{equation}
That is: Under very mild conditions, and even with a number of sub-optimal inequalities being used in the derivation, we have nevertheless been able to see that the total entropy of a star/planet/monster/gravastar/black star/quasi-black hole  is bounded by \emph{half} its area --- this is very close to the holographic bound~\cite{holographic}, which corresponds to $S(\Omega)\leq {1\over4} \A(\partial\Omega)$, and also seems closely related to the generalized second law~\cite{Bekenstein-gsl} and to the Bekenstein bound $S(\Omega) \leq  2\pi E(\Omega) R(\Omega)$~\cite{Bekenstein-bound}. This is also similar in spirit to Srednicki's entanglement entropy~\cite{Srednicki} --- we are bounding the entropy in terms of the visible surface $\partial\Omega$ without looking ``inside'' $\Omega$.  Srednicki's entanglement entropy argument would yield $S(\Omega) \propto \A(\partial\Omega)$ with an unknown and cutoff-dependent proportionality constant.
While our argument provides a precise numerical value for the proportionality constant, ${1\over2}$, unfortunately we have not yet been able to improve the proportionality constant to the ${1\over4}$  one expects based on the holographic bound. On the other hand, the very mildness of the assumptions used in the bound makes it of some independent interest in its own right. 

Fundamentally the reason for this ${1\over2} \leftrightarrow {1\over4}$ mismatch is because we are looking at an uncollapsed distribution of matter, where temperature has its normal interpretation as an intensive variable, and the Euler relation takes the usual uncollapsed form $\rho = T s + \dots$ which for a small element of matter integrates to $M = TS + \dots$  In contrast, once the matter collapses to a black hole then (considering a Schwarzschild black hole for simplicity) $T = 1/(8\pi M)$, so the temperature is no longer an intensive variable. Similarly the Bekenstein entropy $S= {1\over4} \A = 4\pi M^2$ is no longer an extensive variable, and the Euler relation is modified to yield $M = 2 \,TS + \dots$  It is exactly this factor of 2 in the Euler relation for collapsed matter that prevents us from improving our entropy bound for uncollapsed matter to the tighter bound expected for collapsed matter.

As a consistency check consider a static spherically symmetric geometry. Without loss of generality choose coordinates to write the metric in the form
\begin{eqnarray}
\d s^2 &=& - e^{-2\Phi(r)} \left[1-{2m(r)\over r}\right] \d t^2 + {\d r^2\over1-2m(r)/r} 
\nonumber\\
&&+ r^2 \{ \d\theta^2+\sin^2\theta\; \d\phi^2\}.
\end{eqnarray}
Note that with these conventions
\begin{equation}
\sqrt{-g_4} = \sqrt{-g_{00}} \; \sqrt{g_3} = e^{-\Phi} r^2 \sin\theta \to 4\pi r^2 \; e^{-\Phi}.
\end{equation}
We implicitly assume asymptotic flatness, and normalize to $\Phi(\infty)=0$. 
The Killing horizon is defined by the location where $2m(r)/r=1$, that is
\begin{equation}
2m(r_H)=r_H.
\end{equation}
Then it is an old result~\cite{dbh} that at the horizon
\begin{equation}
\kappa_H = {e^{-\Phi_H} (1-2m'_H)\over 2 r_H}.
\end{equation}
By looking at integral curves of the Killing vector, it is easy to see that the 4-acceleration of the FIDOs is 
\begin{eqnarray}
A(r) &=&  \left\{ { m(r) - r m'(r)  \over r^2 \sqrt{1-2m(r)/r}}  - \Phi'(r) \sqrt{1-{2m(r)\over r}} \right\}.
\quad
\end{eqnarray}
A ``red-shifted'' normalized ``generalized surface gravity'' can now be defined for arbitrary $r$ by taking
\begin{equation}
\kappa(r) = {\sqrt{-g_{00}} \; A(r)} = {e^{-\Phi(r)} \; \sqrt{1-2m(r)/r} \; A(r)},
\end{equation}
so
\begin{equation}
\kappa(r) = e^{-\Phi(r)}  \left\{ { m(r) - r m'(r)  \over r^2}  - \Phi'(r) \left[1-{2m(r)\over r}\right] \right\}. \;
\end{equation}
Note that this is now not the surface gravity of the black hole region, but rather the surface gravity of an arbitrary ``virtual sphere'' of radius $r$. Note also that this definition is compatible with that given for the general static case above. (In a different direction this expression is also compatible with that for a time-dependednt spherically symmetric situation as considered in~\cite{Abreu}.) As $r\to r_H$ this tends to the appropriate limit. For all $r$ this has the standard interpertation in terms of the tension in a massless rope supporting a small mass at radius $r$.
A very standard computation now yields~\cite{wormholes}
\begin{eqnarray}
\rho &=& {m'(r)\over4\pi r^2}; \quad
p_r =- {m'(r)\over4\pi r^2} - {\Phi'(r)\over 4\pi r} \left[1-{2m(r)\over r}\right]; 
\nonumber\\
p_t &=& {1\over8\pi } \left\{-{m''(r)\over r}   - \Phi'(r) {1+m(r)/r -3m'(r)\over r}\right.
\nonumber\\
&& \left.  \vphantom{\bigg|} - [\Phi''(r) +(\Phi')^2 ] \; \left[1-{2m(r)\over r}\right] \right\}. 
\end{eqnarray}
By explicit integration one obtains (for \emph{any} $r$)
\begin{equation}
\int_0^r  e^{-\Phi(r)} \,  \{ \rho + p_r + 2 p_t \}\, r^2\, \d r =  r^2 \; \kappa(r).
\end{equation}
That is, in spherical symmetry the partial Tolman mass of a star/planet out to radius $r$ has the very explicit form
\begin{eqnarray}
&& m_T(r) =  r^2 \; \kappa(r) 
\\
&& \quad =  e^{-\Phi(r)}  \left\{ m(r) - r m'(r)   - \Phi'(r) r^2 \left[1-{2m(r)\over r}\right] \right\}.
\nonumber
\end{eqnarray}
Using
\begin{equation}
\kappa(r) = e^{-\Phi(r)} \left\{ {m(r)\over r^2} + 4\pi r p_r \right\},
\end{equation}
it may be advantageous to rephrase this as
\begin{equation}
m_T(r) = e^{-\Phi(r)} \left\{ {m(r)} + 4\pi r^3 p_r \right\}.
\end{equation}
The entropy inequalities still carry through in essentially the same way: For any virtual sphere of radius $r$ we have 
\begin{equation}
S(r) \leq {m_T(r)\over T_\infty} =  {\kappa(r)\; r^2\over T_\infty}.
\end{equation}
By considering FIDOs at radius $r$, the Unruh effect forces 
\begin{equation}
T_\infty \geq {\kappa(r)\over 2\pi},
\end{equation}
so that
\begin{equation}
S(r) \leq 2\pi r^2,
\end{equation}
with this inequality now holding for virtual spheres at arbitrary radii $r$.  The inequality is again sub-optimal, (based on the holographic bound~\cite{holographic} we would have expected $S(r) \leq \pi r^2$), but on the other hand the inequality is extremely robust and easy to derive. 

\def\B{{\cal{B}}}

Now consider the situation where the region $\Omega$ contains a black hole region $\B$ with horizon $\partial\B$.  It makes sense to now define
\begin{equation}
m_T(\Omega) =  m_T(\B) + \int_{\Omega-\B} \sqrt{-g_4} \left\{ \rho + 3 p \right\} \d^3 x,
\end{equation}
where $m_T(\B)$ is the so far undefined Tolman mass to be attributed to the black hole region $\B$. 
Appeal to the flux integral theorem, noting that $\partial(\Omega-\B) = \partial\Omega - \partial\B$,  and using the zeroth law of black hole mechanics to assert that $\kappa(\partial\B)$ is constant on the horizon $\partial\B$~\cite{Mechanix}, to write
\begin{equation}
m_T(\Omega) =  m_T(\B) + {1\over4\pi} \int_{\partial\Omega}  \vec \kappa \cdot  \hat n \; \d \A - {\kappa(\partial\B)\; \A(\partial\B)\over 4\pi}.
\end{equation}
If we now demand that $m_T(\Omega)\to m_{\small ADM}$ at spatial infinity then we uniquely have
\begin{equation}
m_T(\B)  = {\kappa(\partial\B)\; \A(\partial\B)\over 4\pi},
\end{equation}
and
\begin{equation}
m_T(\Omega) =  {1\over4\pi} \int_{\partial\Omega} \vec  \kappa \cdot \hat n \; \d \A .
\end{equation}
That is, with a suitable definition of $m_T(\B)$ we can extend our flux formula for $m_T(\Omega)$ to situations where $\Omega$ contains a black hole region $\B$. 

Finally, consider the total entropy
\begin{equation}
S(\Omega) = S(\B) + \int_{\Omega-\B}  \sqrt{g_3} \; s \; \d^3 x,
\end{equation}
which we have divided into a geometrical entropy associated with the black hole region and a thermodynamic entropy associated with the surrounding matter. 
Again assuming internal equilibrium in the $\Omega-\B$ region, with non-negative pressure $p\geq0$ and non-negative chemical potential $\mu \geq 0$, we obtain the bound
\begin{equation}
S(\Omega) \leq  S(\B) + {m_T(\Omega)-m_T(\B)\over T_\infty}.
\end{equation}
But then our flux theorem gives
\begin{equation}
S(\Omega) \leq  S(\B) + {\int_{\partial\Omega} \vec  \kappa \cdot \hat n \; \d \A -  \kappa(\partial\B)\; \A(\partial\B) \over 4\pi T_\infty},
\end{equation}
implying
\begin{equation}
S(\Omega) \leq  S(\B) + {\bar\kappa(\partial\Omega)   \; \A(\partial\Omega) -  \kappa(\partial\B)\; \A(\partial\B) \over 4\pi T_\infty}.
\end{equation}
But the very fact that we now know $\Omega$ contains a black hole region $\B$ implies that we can see down to the horizon $\partial\B$.
Appealing to the Unruh argument, then at least in normal situations where the surface gravity increases as one moves inwards,  $T_\infty \geq {1\over2\pi} \kappa(\partial\B)$. Therefore
\begin{equation}
S(\Omega) \leq   {\bar\kappa(\partial\Omega)  \over\kappa(\partial\B) }  \; {\A(\partial\Omega)\over2}   +  S(\B) -  {\A(\partial \B)\over2}.
\end{equation}
But then appealing to the ordinary Bekenstein result $S(\B)={1\over4}\A(\partial \B)$, but without any need to invoke the generalized second law or holographic bound, we have
\begin{equation}
S(\Omega) \leq   {\bar\kappa(\partial\Omega)  \over\kappa(\partial\B) }  \; {\A(\partial\Omega)\over2}  .
\end{equation}
This is now a considerably tighter bound on the \emph{total} entropy inside the region $\Omega$, using both information from the surface $\partial\Omega$, plus some information about the black hole region $\B$. 

In summary: We have developed a number of entropy bounds that are very minimalist in the physics ingredients they require.  The Einstein equations are used, mild conditions are placed on pressure and chemical potential, and the Unruh effect is invoked. Even with these rather minimal conditions, quite general and robust bounds can be extracted.

%\vfill
\vskip -20pt
%------------------------------------------------------------------------------------------------------------------------------------------
%------------------------------------------------------------------------------------------------------------------------------------------

%----------------------------------------------------------------------------------------------------------------------------------------

%-----------------------------------------------------------------------------------------------------------------------------------------

%------------------------------------------------------------------------------------------------------------------------------------------
\end{document}